\documentclass[twocolumn]{aastex62}

\received{November 0, 2019}
\revised{February 13, 2020}
\accepted{February 13, 2020}
\submitjournal{ApJ}

\shorttitle{Expansion of Nova Shells}
\shortauthors{E. Santamar\'\i a et al.}

\begin{document}

\title{Angular Expansion of Nova Shells\footnote{Released on November, 8th, 2019}}

\correspondingauthor{Edgar Santamar\'\i a}
\email{edgar.santamaria.dom@gmail.com, mar@iaa.es}

\author[0000-0003-4946-0414]{E.\ Santamar\'\i a}
\affil{CUCEI, Universidad de Guadalajara, Blvd. Marcelino Garc\'\i a Barrag\'an 1421, 44430, Guadalajara, Jalisco, Mexico}
\affiliation{Instituto de Astronom\'\i a y Meteorolog\'\i a, Dpto.\ de F\'\i sica,
CUCEI, Av.\ Vallarta 2602, 44130, Guadalajara, Jalisco, Mexico}

\author[0000-0002-7759-106X]{M.A.\ Guerrero}
\affiliation{Instituto de Astrof\'\i sica de Andaluc\'\i a, IAA-CSIC, Glorieta de la
Astronom\'\i a s/n, 18008, Granada, Spain}

\author[0000-0003-2653-4417]{G.\ Ramos-Larios}
\affiliation{CUCEI, Universidad de Guadalajara, Blvd. Marcelino Garc\'\i a Barrag\'an 1421, 44430, Guadalajara, Jalisco, Mexico}
\affiliation{Instituto de Astronom\'\i a y Meteorolog\'\i a, Dpto.\ de F\'\i sica,
CUCEI, Av.\ Vallarta 2602, 44130, Guadalajara, Jalisco, Mexico}

\author[0000-0002-5406-0813]{J.A.\,Toal\'{a}}
\affiliation{Instituto de Radioastronom\'{i}a y Astrof\'{i}sica (IRyA), UNAM Campus Morelia,
Apartado postal 3-72, 58090 Morelia, Michoac\'{a}n, Mexico}

\author[0000-0003-0242-0044]{L.\ Sabin}
\affiliation{Instituto de Astronom\'\i a, Universidad Nacional Aut\'onoma de M\'exico,
Apdo.\ Postal 877, C.P. 22860, Ensenada, B.C., Mexico}

\author[0000-0002-9284-4504]{G.\ Rubio}
\affiliation{CUCEI, Universidad de Guadalajara, Blvd. Marcelino Garc\'\i a Barrag\'an 1421, 44430, Guadalajara, Jalisco, Mexico}
\affiliation{Instituto de Astronom\'\i a y Meteorolog\'\i a, Dpto.\ de F\'\i sica,
CUCEI, Av.\ Vallarta 2602, 44130, Guadalajara, Jalisco, Mexico}

\author[0000-0003-4282-8108]{J.A.\ Quino-Mendoza}
\affiliation{CUCEI, Universidad de Guadalajara, Blvd. Marcelino Garc\'\i a Barrag\'an 1421, 44430, Guadalajara, Jalisco, Mexico}
\affiliation{Instituto de Astronom\'\i a y Meteorolog\'\i a, Dpto.\ de F\'\i sica,
CUCEI, Av.\ Vallarta 2602, 44130, Guadalajara, Jalisco, Mexico}

\begin{abstract}

Nova shells can provide us with important information on their distance,
their interactions with the circumstellar and interstellar media, and the
evolution in morphology of the ejecta.  
We have obtained narrow-band images of a sample of five nova shells,
namely DQ\,Her, FH\,Ser, T\,Aur, V476\,Cyg, and V533\,Her, with ages
in the range from 50 to 130 years.  
These images have been compared with suitable available
archival images to derive their angular expansion rates.  
We find that all the nova shells in our sample are still in
the free expansion phase, which can be expected, as the mass
of the ejecta is 7-45 times larger than the mass of the
swept-up circumstellar medium.
The nova shells will keep expanding freely for time periods up to a
few hundred years, reducing their time dispersal into the interstellar
medium
 
\end{abstract}


\section{Introduction} \label{sec:intro}

Novae are the result of the interaction of stars in close binary systems, 
where a white dwarf (WD) accretes H-rich material from a companion, typically 
a giant or sub-giant low-mass star \citep{2008B}. 
When the accreted material reaches a critical mass, a thermonuclear runaway 
(TNR) occurs. Temperatures can reach values of $\sim$1--4$\times$10$^{8}$ K in a few 
seconds \citep{2016PASP..128e1001S} and up to $\sim$2$\times$10$^{-4}$ 
M$_{\odot}$ \citep{1998PASP..110....3G} of highly processed material is 
ejected at velocities $\sim$1000 km s$^{-1}$ \citep{2010AN....331..160B} 
in a classical nova (CN) event.  
With time, the nova remnants will expand and mix into the interstellar 
medium (ISM).

\begin{table*}\centering
\setlength{\columnwidth}{0.1\columnwidth}
\setlength{\tabcolsep}{1.0\tabcolsep}
\caption{Sample of Novae \label{tab:nov}}
\begin{tabular}{lrlllcccc}
\hline
\hline

\multicolumn{1}{c}{Object} &
\multicolumn{1}{c}{$l,b$} & 
\multicolumn{1}{c}{Nova Type} &
\multicolumn{1}{c}{Outburst Date} & 
\multicolumn{1}{c}{Distance} & 
\multicolumn{1}{c}{$z$} &
\multicolumn{1}{c}{Angular Size} & 
\multicolumn{1}{c}{$v_{\rm exp}^{\rm sp}$} & 
\multicolumn{1}{c}{References} \\ 

\multicolumn{1}{c}{} &
\multicolumn{1}{c}{} & 
\multicolumn{1}{c}{} & 
\multicolumn{1}{c}{} & 
\multicolumn{1}{c}{} & 
\multicolumn{1}{c}{} & 
\multicolumn{1}{c}{major$\times$minor} & 
\multicolumn{1}{c}{major$\times$minor} & 
\multicolumn{1}{c}{} \\ 

\multicolumn{1}{c}{} &
\multicolumn{1}{c}{($^\circ$)} & 
\multicolumn{1}{c}{} & 
\multicolumn{1}{c}{} & 
\multicolumn{1}{c}{(pc)} & 
\multicolumn{1}{c}{(pc)} & 
\multicolumn{1}{c}{($^{\prime\prime}$)} &
\multicolumn{1}{c}{(km s$^{-1}$)} & 
\multicolumn{1}{c}{} \\ 
\hline

T\,Aur    & 177.14$-$1.70  & fast      & 1891 Dec & ~~880$^{+50}_{-35}$  & ~~25 & 25.4$\times$18.6 &       655      & 1 \\
V476\,Cyg &  87.37$+$12.42 & very fast & 1920 Aug & ~~670$^{+110}_{-50}$ &  145 & 14.6$\times$13.4 &       725      & 2 \\
DQ\,Her   &  73.15$+$26.44 & slow      & 1934 Dec & ~~501$^{+6}_{-6}$    & 220 &  32.0$\times$24.2 &       370      & 3 \\
V533\,Her &  69.19$+$24.27 & slow      & 1963 Feb &  1200$^{+50}_{-40}$  &  495 & 16.8$\times$15.2 &       850      & 4 \\
FH\,Ser   &  32.91$+$5.79  & slow      & 1970 Feb &  1060$^{+110}_{-70}$ & 105 &  12.4$\times$10.6 & 490$\times$385 & 4 \\

\hline
\end{tabular}
\vspace{0.15cm}
\tablecomments{
Distances were obtained from \emph{Gaia} DR2 \citep{2018MNRAS.481.3033S}. 
References for spectroscopic expansion velocities.--
(1) \citet{1983ApJ...268..689C},
(2) \citet{DUER1987},
(3) \citet{VOR2007},
(4) \citet{GIL200}.
}
\end{table*}

The morphology and expansion of a nova remnant depend on the details of the
nova event, but also on the interactions of the ejecta with the stellar
companion and the pre-existing circumstellar material, which may consists of
an accretion disk and a common envelope.  
A typical CN outburst includes an initial slow (500--2000 km~s$^{-1}$) 
wind followed by a longer phase with a faster (1000--4000 km~s$^{-1}$) 
wind \citep{BODE1989}.  
The interaction of these two winds forms a double shock structure, 
with the fast wind passing through the slow one until it dissipates 
and cools adiabatically as it expands \citep{OBRIEN1994}.  

Additionally, the interaction of the ejecta with the binary companion and 
material in a common envelope has effects in the asphericity of the nova
shell \citep{LIVIO1990} and dynamics of the ejecta \citep{SHANKAR1991}.
The effects of these interactions vary among novae of different speed
class \citep{LLOYD1997}, which are basically associated to the different
time-scales of the slow and fast wind phases, providing an interpretation
for the larger asphericities of the remnants of slow novae with respect to
those of fast novae \citep{1995MNRAS.276..353S}.  
Finally, the WD rotation may also feed the ejecta with 
angular momentum, which can produce noticeable effects 
on the structure of nova shells \citep{PORTER1998}.

The late expansion of nova shells can help us gain insights into the
plasma physics and shock phenomena associated
with the blast produced by the interaction of hydrogen-poor, metal-rich
ejecta with the circumstellar environment and to investigate the
ingestion of this ejecta by the ISM.
The complete dynamical evolution of a nova occurs in time scales comparable
to that of human life, and thus it provides a first class comparison to
investigate the much slower evolution of planetary nebulae (PNe) or the
processes involved in the evolution of the much rare supernova remnants
(SNR).
The time scale for a nova dispersal is an important parameter
to assess the duration of the different stages of hibernation
between a CN eruption and their parents cataclysmic variables
\citep{SHARA2017}.

Very little attention has been paid to the late expansion of nova shells,
however.  
\citet{DUER1987} conducted a heroic investigation of the angular expansion 
of nova shells using images of limited quality and concluded that they have
mean half-time of 75 years, noting that this deceleration is most noticeable
for novae with higher expansion velocities.  
Since then, very few detailed studies of the angular expansion of nova
shells have been carried out, including those of GK\,Per, perhaps the
most studied nova shell \citep{LIIM2012,TAK2015,HAR2016}, DQ\,Her 
\citep{HER1992,VOR2007,HARR2013}, FH\,Ser \citep{VAL1997,ESEN1997}, 
and recently IPHASX\,J210204.7$+$471015 \citep{SAN2019}.
The multiple knots in GK\,Per expand isotropically at an angular velocity
of 0\farcs3-0\farcs5 yr$^{-1}$, which has been kept unchanged since their
ejection a century ago \citep{Shara2012b,LIIM2012}.  
This is somehow surprising, because detailed analyses of individual
knots reveal the notable interaction with each other and with the ISM
\citep{HAR2016}.
On the other hand, a noticeable deceleration of a bow-shock
component of the nova IPHASX\,J210204.7$+$471015 has been 
recently reported \citep{SAN2019}.

The lack of agreement between these results most likely implies
that the expansion of a nova shell depends on the details of the
nova outburst and the local properties of the ISM.
The availability of high-quality archival images of nova
shells allows a precise investigation of the expansion of
a meaningful sample of sources.  
Using the sample of images presented by \cite{1995MNRAS.276..353S},
we have selected five nova shells with multi-epoch high-quality
images, namely DQ\,Her, FH\,Ser, T\,Aur, V476\,Cyg, and V533\,Her, 
to carry out a pilot study of the investigation of the angular
expansion of nova shells.  
Basic information on these novae, including their Galactic coordinates,
type, outburst date, distance
\citep[as adapted from][]{2018MNRAS.481.3033S},
height over the Galactic Plane, and
expansion velocity derived from spectroscopic observations,  
is compiled in Table~\ref{tab:nov}.  
The latter is provided for the major and minor axes of FH\,Ser.

\begin{figure*}
\centering
\includegraphics[width=0.95\textwidth]{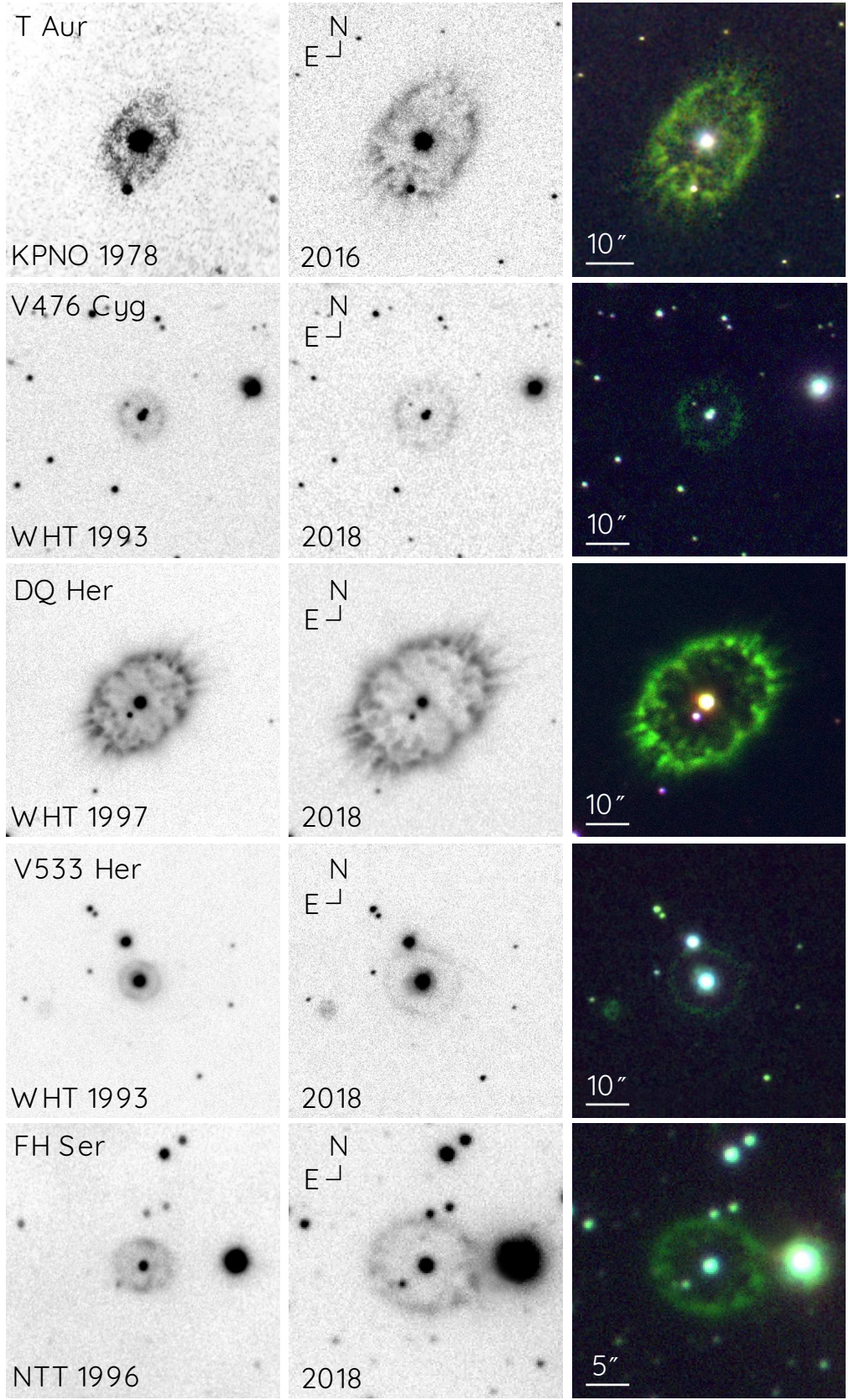}
\caption{
From top to bottom, multi-epoch H$\alpha$ images (left and middle 
panels) of T\,Aur, V476\,Cyg, DQ\,Her, V533\,Her, and FH\,Ser 
(see Table~\ref{tab:obs} for details) and \emph{RGB} composite 
colour pictures (right panel) combining NOT ALFOSC images in the 
broadband $g'$ SDSS $\lambda$4800 (blue) and narrowband H$\alpha$ 
$\lambda$6563 (green) and [N~{\sc ii}] $\lambda$6583 (red) filters, 
but for V476\,Cyg, whose colour picture was obtained using an $r'$ 
SDSS $\lambda$6180 filter for the red colour.  
\label{fig:fig1}}
\end{figure*}

\section{Imaging} \label{sec:obs}

\subsection{Contemporary Imaging} \label{sec:own_img}

Present day (2016-2019) images of the nova shells in Table~\ref{tab:nov}
were obtained using the Alhambra Faint Object Spectrograph and Camera
(ALFOSC) at the 2.5m Nordic Optical Telescope (NOT) of the Roque de los
Muchachos Observatory (ORM) in La Palma, Spain. 
The E2V 231-42 2k$\times$2k CCD was used with pixel size 15.0 $\mu$m,
providing a plate scale of 0\farcs211 pix$^{-1}$ and a field of view
(FoV) of 7\farcm2 arcmin.
The images used to investigate the angular expansion of these novae were
obtained through H$\alpha$ filters with FWHM of 33 \AA\ for the 2016 run
of T\,Aur and 13 \AA\ for the others.  
Total exposure times and spatial resolutions, as determined from 
the FWHM of field stars, are listed in Table~\ref{tab:obs}.

Images were also acquired in other filters as described in the caption
of Figure~\ref{fig:fig1} to obtain colour-composite pictures of these
novae.  
All images were processed using standard {\sc iraf} routines.

\subsection{Archival Imaging} \label{sec:arc_img}

Archival CCD images of the novae in Table~\ref{tab:nov} have been 
obtained using different telescopes and instruments as listed in 
Table~\ref{tab:obs}.
The images were downloaded from
the European Southern Observatory (ESO) Science Archive Facility
the Isaac Newton Group (ING) data archive
and the Mikulski Archive for Space Telescopes (MAST) and Hubble
Legacy Archive (HLA) at the Space Telescope Science Institute 
The ESO images were obtained using SUper Seeing Instrument (SUSI)
and ESO Faint Object Spectrograph and Camera 2 (EFOSC2) at the 3.5m 
New Technology Telescope (NTT) of the ESO's La Silla Observatory.
The ING images were acquired using the Auxiliary-port CAMera (ACAM)
of the 4.2m William Herschel Telescope (WHT) and Jacobus Kapteyn 
Telescope (JKT).

The \emph{HST} images were obtained using the Wide Field and Planetary
Camera 2 \citep[WFPC2 Instrument Handbook,][]{2009wfpc.rept....4B} under
programs ID 6770 (PI O'Brien) and 6060 (PI Shara).
The filters, exposure times, pixel scales, and spatial resolutions
of these images are listed in Table~\ref{tab:obs}.

\begin{table*}[t!]
\centering
\setlength{\columnwidth}{0.1\columnwidth}
\setlength{\tabcolsep}{1.0\tabcolsep}
\caption{Details of the Imaging Observations \label{tab:obs}}
\begin{tabular}{llllrcc}
\hline
\hline

\multicolumn{1}{c}{Object} &
\multicolumn{1}{c}{Date} &
\multicolumn{1}{c}{Telescope and} & 
\multicolumn{1}{c}{Filter} & 
\multicolumn{1}{c}{Exposure} & 
\multicolumn{1}{c}{Pixel} &
\multicolumn{1}{c}{Spatial} \\ 

\multicolumn{1}{c}{} &
\multicolumn{1}{c}{} & 
\multicolumn{1}{c}{Instrument} & 
\multicolumn{1}{c}{} & 
\multicolumn{1}{c}{Time} & 
\multicolumn{1}{c}{Scale} &
\multicolumn{1}{c}{Resolution} \\ 

\multicolumn{1}{c}{} &
\multicolumn{1}{c}{} & 
\multicolumn{1}{c}{} & 
\multicolumn{1}{c}{} &
\multicolumn{1}{c}{(s)} & 
\multicolumn{1}{c}{($^{\prime\prime}$ pix$^{-1}$)} &
\multicolumn{1}{c}{($^{\prime\prime}$)} \\ 
\hline

T\,Aur    & 1956 Dec    & PO \& 103aE            & Red       & $\dots$~~~~     & 1.7 &  $\dots$   \\
          & 1978 Mar    & KPNO \& ISIT           & H$\alpha$ & 1800~~~~     & $\dots$ & $\dots$ \\
          & 1989 Nov 22 & POSS2                  & Red RG610 & 4800~~~~     & 1.0  & $\dots$ \\
 	  & 1998 Nov 2  & \textit{HST}  \& WFPC2 & F656N     & 5400~~~~ & 0.05 & 0.2\\
 	  & 2016 Nov 28 & NOT \& ALFOSC          & NOT \#21 H$\alpha$ & 1800~~~~ & 0.21 & 0.6\\
  	  & 2018 Jan 03 & NTT \& EFOSC2          & H$\alpha$ & 1440~~~~ & 0.12 & 0.5\\
  	  & 2019 Oct 11 & NOT \& ALFOSC          & OSN H01 H$\alpha$ & 2700~~~~ & 0.21 & 0.9\\
V476\,Cyg & 1944 Jan-Jun & MW \& 100-inch        & H$\alpha$ &     $\dots$~~~~    &   $\dots$   &   $\dots$  \\
 	  & 1993 Sep 12 & WHT \& Aux.\ Port      & H$\alpha$ 6569 & 900~~~~ & 0.25 & 1.2\\
	  & 2018 Jun 08 & NOT \& ALFOSC          & OSN H01 H$\alpha$ & 2700~~~~ & 0.21 & 0.7\\
DQ\,Her   & 1977 May 15 & BokT \& ITT 40mm       & H$\alpha$ &    $\dots$~~~~      &   $\dots$   &   $\dots$  \\
          & 1993 Jul 31 & JKT \& AGBX            & H$\alpha$ & 7200~~~~ & 0.33 & 2.0 \\
          & 1995 Sep 04 & \emph{HST} \& WFPC2    & F656N     & 2000~~~~ & 0.05 & 0.1 \\
          & 1997 Oct 25 & WHT \& Aux.\ Port      & H$\alpha$ 656     & 1200~~~~ & 0.11 & 0.4 \\
 	  & 2012 Aug 15 & WHT \& ACAM            & T6565    & 40~~~~ & 0.25 & 0.7\\
  	  & 2017 May 27 & NOT \& ALFOSC          & OSN H01 H$\alpha$ & 2700~~~~  & 0.21 & 0.8\\
  	  & 2018 Jun 05 & NOT \& ALFOSC          & OSN H01 H$\alpha$ & 2700~~~~  & 0.21 & 0.8\\ 
V533\,Her & 1993 Sep 11 & WHT \& Aux.\ Port      & H$\alpha$ 6569      & 1800~~~~ &   0.25   & 1.0 \\
 	  & 1997 Sep 03 & \emph{HST} \& WFPC2    & F656N & 2600~~~~  & 0.05 & 0.2\\
 	  & 2018 Jun 06 & NOT \& ALFOSC          & OSN H01 H$\alpha$ & 4800~~~~  & 0.21 & 0.6\\
FH\,Ser	  & 1996 Mar 18 & NTT \& SUSI            & H$\alpha$ &  720~~~~ & 0.13 & 0.9 \\
  	  & 1997 May 11 & \textit{HST} \& WFPC2  & F656N & 4800~~~~  & 0.05 & 0.1\\
  	  & 2017 May 29 & NOT \& ALFOSC          & OSN H01 H$\alpha$ & 2700~~~~  & 0.21 & 0.6\\
  	  & 2018 Jun 06 & NOT \& ALFOSC          & OSN H01 H$\alpha$ & 3600~~~~  & 0.21 & 0.7\\
\hline
\end{tabular}
\vspace{0.15cm}
\end{table*}

Ancient images were acquired using photographic plates.
The 1956 image of T\,Aur was taken by Walter Baade at the Palomar
Observatory 
\citep{1970Ap&SS...6..183M}, whereas that of 1978 was obtained at Kitt 
Peak National Observatory (KPNO) using the 4m telescope
\citep{1980ApJ...237...55G}.
The oldest image of V\,476 Cyg was acquired in 1944 at Mount Wilson
Observatory 
\citep[for more details, see][]{1944MWOAR..16....1A,1970IAUS...39..281B}.
The 1977 image of DQ\,Her was obtained with the Steward Observatory
2.3m telescope using an ITT 40 mm tube 
\citep{1978ApJ...224..171W}.

\begin{figure}
\centering
\includegraphics[bb=20 40 635 505,width=0.48\textwidth]{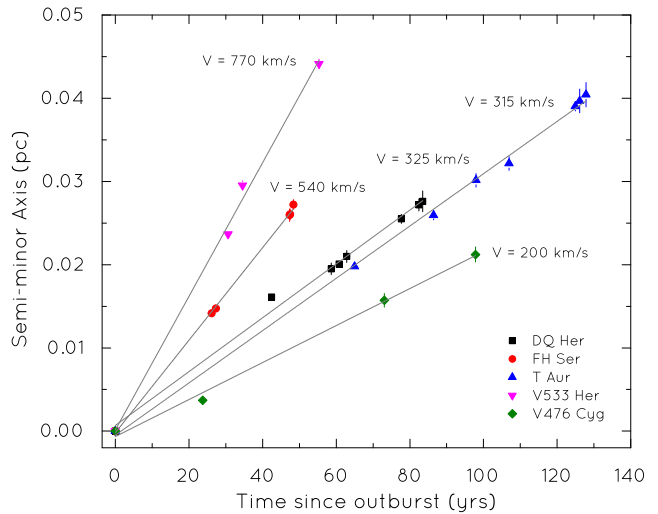}
\caption{
Expansion of the semi-minor axis of nova shells.  
The angular size of the semi-minor axis measured in the images
has been converted into linear size in pc using the \emph{Gaia}
DR2 distance to each nova as listed in Table~\ref{tab:nov},
whereas the epoch of each measurement is referred to the time
since the nova outburst.  
The error bars correspond to the dispersion of the individual values
obtained for each epoch, which is smaller than the symbol size in a
few cases.
The expansion of all novae is consistent with free expansion.
The slope of the linear fits has been converted to
expansion velocity in the common units of km~s$^{-1}$.  
\label{fig:fig2}}
\end{figure}

\section{Results} \label{sec:res}

We present in the left and middle columns of Figure~\ref{fig:fig1}
archival and present day images of the nova shells in our sample,
respectively. 
These images reveal all nova shells in our sample to have
elliptical morphologies with different degree of ellipticity.
T\,Aur and DQ\,Her have similar knotty morphologies, with
cometary knots showing remarkable tails mainly along the
major axis.
V476\,Cyg also seems to have a broken, clumpy morphology,
whereas the shells of FH\,Ser and V533\,Her have smoother
appearance.  
The comparison of present day images (Fig.~\ref{fig:fig1} middle column)
with representative archival images (Fig.~\ref{fig:fig1} left column)
unveils clear expansion patterns. 
A careful examination of multi-epoch images also discloses
small-scale morphological variations, including changes in
the size and distribution of clumps.
A detailed study is deferred to a subsequent work
(Santamar\'\i a et al., in preparation).

To investigate and quantify the expansion of these nova shells,
radial spatial profiles across individual discrete features have
been extracted from the images at the different epochs listed in
Table~\ref{tab:obs}.
The distance of these features to the central star has been
determined by measuring their position using Gaussian fits.
The angular sizes along different directions have then been
normalized to the minor axis using an elliptical fit to the 
shape of the nova shell, and an averaged value for the size
of the minor axis and its 1$\sigma$ dispersion derived for
each epoch.
These are shown in Figure~\ref{fig:fig2}, together with 
linear fits for all
the nova shells in our sample, 
where the time is computed from the nova outburst date and the
angular size of the semi-minor axis has been converted to linear
size using the nova distance. 
The increase of the size of these nova shells with time can be described by
linear fits (Figure~\ref{fig:fig2}) with correlation coefficients $\geq$0.98,
implying $t$-test significance probabilities $\geq$98\% for V533\,Her and
V476\,Cyg, $>$99\% for FH\,Ser, and $>$99.9\% for DQ\,Her and T\,Aur.
The slope of these fits correspond to the angular expansion rates along
the minor axis of these nova shells (column 2 of Table~\ref{tab:nov}).

We note that the angular expansion rates derived from these fits
are consistent with those previously reported.
 
\cite{VOR2007} derived angular expansion rates of 0\farcs205$\pm$0\farcs014 yr$^{-1}$
and 0\farcs165$\pm$0\farcs012 yr$^{-1}$ along the major and minor axes of DQ\,Her, respectively, whereas angular expansion rates of
0\farcs128~yr$^{-1}$ \citep{SD1987}, 
0\farcs136~yr$^{-1}$ \citep{DUER1992}, and 
0\farcs104--0\farcs146~yr$^{-1}$ \citep{VAL1997}
have been reported for FH\,Ser. 
Similarly, \citet{Harvey2018} provides angular expansion rates $\approx$0\farcs12~yr$^{-1}$ and $\approx$0\farcs088~yr$^{-1}$ along the major and minor axes of T\,Aur, respectively, and $\approx$0\farcs075~yr$^{-1}$ for V476\,Cyg.

\begin{table*}\centering
\setlength{\columnwidth}{0.1\columnwidth}
\setlength{\tabcolsep}{1.0\tabcolsep}
\caption{Flux and mass of the nova shells \label{tab:mass}}
\begin{tabular}{lcccccc}
\hline
\hline

\multicolumn{1}{c}{Object} &
\multicolumn{1}{c}{Angular Expansion} & 
\multicolumn{1}{c}{$\bar{v}_{\rm exp}$} & 
\multicolumn{1}{c}{$M_{\rm swept} \times (\frac{n_{\rm ISM}}{1~{\rm cm}^{-3}})$} & 
\multicolumn{1}{c}{$M_{\rm shell} \times (\frac{\epsilon}{0.1})^{0.5}$} &
\multicolumn{1}{c}{$F_{{\rm H}\alpha}$} & 
\multicolumn{1}{c}{$E_{\rm kin}$} \\
\multicolumn{1}{c}{} &
\multicolumn{1}{c}{($^{\prime\prime}$~yr$^{-1}$)} &
\multicolumn{1}{c}{(km s$^{-1}$)} & 
\multicolumn{1}{c}{($M_{\rm \odot}$)} &
\multicolumn{1}{c}{($M_{\rm \odot}$)} & 
\multicolumn{1}{c}{(erg cm$^{-2}$ s$^{-1}$)} & 
\multicolumn{1}{c}{(erg)} \\

\hline

T\,Aur    & (0.097$\pm$0.004)$\times$(0.072$\pm$0.001) & (410$\pm$40)$\times$(315$\pm$22) & 1.3$\times$10$^{-5}$ & 3.6$\times$10$^{-4}$ & 2$\times$10$^{-13}$ & 4.2$\times$10$^{44}$ \\ 
V476\,Cyg & (0.073$\pm$0.008)$\times$(0.067$\pm$0.007) & (230$\pm$60)$\times$(200$\pm$50) & 1.7$\times$10$^{-6}$ & 2.2$\times$10$^{-5}$ & 1$\times$10$^{-14}$ & 1.1$\times$10$^{43}$ \\ 
DQ\,Her   & (0.188$\pm$0.008)$\times$(0.139$\pm$0.005) & (460$\pm$25)$\times$(325$\pm$16) & 5.2$\times$10$^{-6}$ & 2.3$\times$10$^{-4}$ & 7$\times$10$^{-13}$ & 3.3$\times$10$^{44}$ \\
V533\,Her & (0.152$\pm$0.006)$\times$(0.139$\pm$0.007) & (850$\pm$70)$\times$(770$\pm$70) & 1.3$\times$10$^{-5}$ & 9.0$\times$10$^{-5}$ & 7$\times$10$^{-15}$ & 5.9$\times$10$^{44}$ \\ 
FH\,Ser   & (0.125$\pm$0.002)$\times$(0.109$\pm$0.002) & (630$\pm$80)$\times$(540$\pm$70) & 3.5$\times$10$^{-6}$ & 1.4$\times$10$^{-4}$ & 9$\times$10$^{-14}$ & 4.7$\times$10$^{44}$ \\

\hline
\end{tabular}
\vspace{0.15cm}
\end{table*}

\section{Discussion}

The main result from the investigation of the angular expansion of
this sample of nova shells is their linear expansion with time
(Fig.~\ref{fig:fig2}).  
This linear increase of size with time implies a free expansion, where the
initial velocity of the ejecta remains the same since the nova event with
no sign of deceleration.
Thus, we should expect the expansion velocity of a nova shell derived
from its angular expansion rate and distance ($\bar{v}_{\rm exp}$,
column 3 of Table~\ref{tab:mass}), which is the averaged expansion
velocity of the ejecta since the nova outburst
projected on the plane of the sky,
to be consistent with
the expansion velocity derived from spectroscopic observations
($v_{\rm exp}^{\rm sp}$, column 8 of Table~\ref{tab:nov}), which is the
expansion velocity along the line of sight
at the time of the observation\footnote{
  Spatiokinematic models of nova shells found them to be described as
  prolate ellipsoid with axial ratios $\leq$1.3 \citep[e.g.,
  FH\,Ser][]{GIL200}.
  Since the angular expansion along the minor axis probes the equatorial
  expansion of such a prolate ellipsoid, the spectroscopic velocity can
  be expected to be at most 1.3 times larger than the latter in the most
  favorable case of pole-on ellipsoids.}.
These two expansion velocities are found to be in
excellent agreement for DQ\,Her and V533\,Her, and
within the uncertainties for FH\,Ser, whose
$v_{\rm exp}^{\rm sp}$ have been derived from
spatiokinematic models.
Remarkable discrepancies are found, however, for T\,Aur
and V476\,Cyg, whose spectroscopic observations are of
low quality \citep{1983ApJ...268..689C,DUER1987}.

  An orientation of the major axis of these nova shells 
  close to the line of sight cannot explain the much larger spectroscopic 
  velocities of T\,Aur and V476\,Cyg than the expansion velocities on the plane 
  of the sky along the minor axis. We note that recent high-dispersion spectra 
  of T\,Aur imply expansion velocities similar to those found here \citep{Harvey2018}. 

The linear expansion with time of the nova shells in this sample strengthens 
the idea that the ejecta has kept expanding at
its initial velocity since the nova event. 
This result confirms previous results presented for T\,Aur and V476\,Cyg, but also for V1500\,Cyg and V4362\,Sgr \citep{Harvey2018}. 
Apparently, the circumstellar medium around these novae has not
been able to slow down their expansion, which can be expected if
the mass of the ISM material swept up by the nova shell is much
smaller than the mass of the nova ejecta.
This can be tested by computing their values.  
Assuming an ISM density\footnote{
  The averaged density of the ISM along the path
  towards these novae is in the range from 0.1 to 
  2 cm$^{-3}$ \citep{HI4PI2016}.
}

$n_{\rm ISM}$=1 cm$^{-3}$,
the volume evacuated by the nova shell implies
swept-up masses of 10$^{-6}$--10$^{-5}$ $M_\odot$
(column 4 in Table~\ref{tab:mass}).  
These can be compared with the nova masses
2$\times$10$^{-5}$--3$\times$10$^{-4}$ $M_\odot$
(column 5 in Table~\ref{tab:mass}).
The latter have been derived following \citet{1970Ap&SS...6..183M},
using the unabsorbed H$\alpha$ fluxes listed in column 6
and assuming a filling factor $\epsilon$=0.1.  
The H$\alpha$ fluxes are computed from our H$\alpha$ narrow-band images,
using intermediate-dispersion flux-calibrated spectra of DQ\,Her to derive
a count-to-flux conversion factor and corrected for absorption using the extinction values given by \citet{1995MNRAS.276..353S} for T\,Aur and V476\,Cyg, \citet{SG2013} for DQ\,Her and V533\,Her, and \citet{GIL200} for FH\,Ser.
The masses of the nova ejecta are indeed much greater
than the masses of the swept up ISM, by factors from
7 to 45, which is consistent with their free expansion.
At their present expansion rates, the free expansion can be expected to last
from $\simeq$100 yr for V533\,Her up to $\simeq$400 yr for T\,Aur until the 
time when the swept up ISM mass equals that of the nova ejecta.
Since the nova ejecta is not slowed down, it reduces
the time for dispersal of nova shells into the ISM.  

The free expansion is supported by the large kinetic energy 
($\frac{1}{2}\,M_{\rm shell}\,v_{\rm exp}^2$) of the nova shells,
which have been computed adopting a weighted expansion velocity
among the polar and equatorial velocities.
The kinetic energies listed in column 7 of Table~\ref{tab:mass}
are in the range of a few times 10$^{44}$ erg, but for V476\,Cyg,
which is 1$\times$10$^{43}$ erg.

The free expansion of the nova shells in our sample is in sharp contrast
with the conclusions drawn by \citet{DUER1987}, who proposed that the
expansion velocity of a nova shell reduces to half every 75 years.
This is particularly shocking for DQ\,Her and V476\,Cyg (this work),
and GK\,Per \citep{Shara2012b,LIIM2012}, which were proposed to have
deceleration half-times of 67, 117, and 58 years, respectively.
The free expansion of nova shells applies to different nebular
morphologies, from the smooth elliptical morphology of FH\,Ser,
V476\,Cyg, and V533\,Her, the mildly broken elliptical structure
of T\,Aur and DQ\,Her, and the knotty morphology of GK\,Per.
Only the faint bow-shock structural component of
IPHASX\,J210204.7$+$471015 seems to have experienced
a notable braking in its interaction with the ISM
\citep{2018ApJ...857...80G,SAN2019}.

\section{Summary and Conclusions}

The comparison between multi-epoch suitable broadband and
narrowband images of the nova shells DQ\,Her, FH\,Ser,
T\,Aur, V476\,Cyg, and V533\,Her has been used to derive
their angular expansion rates.  
This is found to be unchanged since the nova event,
i.e., the nova shells in this sample are still in
a free expansion phase.
This can be expected, as the mass of the ejecta is 7-45 times
larger than the mass of the swept-up circumstellar medium.

The images of the nova shells in our sample cover a 
time lapse since the nova event from 20 to 130 yrs.

Given the large ratio between the mass of the ejecta and that of the swept-up
circumstellar medium, the free expansion phase can be expected to last for a few hundred years, during most (if not all) their whole visible phase.  

\acknowledgments

E.S.\, G.R.\ and J.A.Q.M.\ acknowledges support from CONACyT and Universidad de Guadalajara.
M.A.G.\ and E.S.\ acknowledge financial support by grants AYA~2014-57280-P 
and PGC~2018-102184-B-I00, co-funded with FEDER funds.  
M.A.G.\ acknowledges support from the State Agency for Research of the 
Spanish MCIU through the ``Center of Excellence Severo Ochoa'' award for 
the Instituto de Astrof\'\i sica de Andaluc\'\i a (SEV-2017-0709).
E.S.\ acknowledges the hospitality of the IAA during a short-term visit.
G.R.L.\ acknowledges support from CONACyT and PRODEP (Mexico).
L.S.\ acknowledges support from UNAM DGAPA PAPIIT project IN101819.
M.A.G.\ and J.A.T.\ acknowledge support from the UNAM DGAPA PAPIIT project IA 100318.
We appreciate the valuable comments of the referee, Dr Nye Evans.
Finally, we thank Alessandro Ederoclite for useful discussion and comments.
The data presented here were obtained in part with ALFOSC, which is provided by the Instituto de Astrof\'\i sica de Andaluc\'\i a (IAA) under a joint agreement with the University of Copenhagen and NOTSA.
This research made use of {\sc iraf}, distributed by the National Optical Astronomy Observatory, which is operated by the Association of Universities for
Research in Astronomy (AURA) under a cooperative agreement
with the National Science Foundation. 
We acknowledge the use of The ESO Science Archive Facility developed in partnership with the Space Telescope European Coordinating Facility (ST-ECF).
Also, the ING archive, maintained as part of the CASU Astronomical Data Centre at the Institute of Astronomy, Cambridge and finally, the STScI, operated by the Association of Universities for research in Astronomy, Inc., under NASA contract NAS5-26555.

\vspace{5mm}
\facilities{HST(STIS), NOT:2.5m}

\software{astropy \citep{2013A&A...558A..33A},  
          {\sc iraf} \citep{Tody1986}
          }

\end{document}